\documentstyle[prb,aps,twocolumn]{revtex}
\large
\begin{document}
\draft
\twocolumn[\hsize\textwidth\columnwidth\hsize\csname
@twocolumnfalse\endcsname

\title{Magnetic Resonance of Spin Clusters and Triplet Excitations in
a Spin-Peierls Magnet with Impurities}

{\it to be published in JETP in July 2001}
\vspace{10mm}

\author{V.~N.~Glazkov, A.~I.~Smirnov}
\address{P.~L.~Kapitza  Institute for  Physical  Problems  RAS,
117334 Moscow, Russia}

\author{ R. M. Eremina}
\address{
 Zavoiskii Institute of Engineering Physics, Kazan, 420029 Russia}

\author{ G. Dhalenne, A. Revcolevschi }
\address{
 Laboratoire de Physico-Chimie de l'Etat Solide, Universite@ Paris-Sud,
Orsay Cedex, 91405 France }

\date{\today}
\maketitle

\begin{abstract}
\widetext \leftskip 54.8pt \rightskip 54.8pt The magnetic
resonance spectrum of spin clusters formed in spin-Peierls magnets
in the vicinity of impurity ions was investigated. The observed
temperature dependences of the effective $g$-factor and the
linewidth of the electron spin resonance (ESR) in crystals of
Cu$_{1-x}$Ni$_x$GeO$_3$ are described in the model of the exchange
narrowing of the two-component spectrum with one component
ascribed to spin clusters and exhibiting an anomalous value of the
$g$-factor and the other related to triplet excitations. An
estimation of the size of the suppressed dimerization region
around the impurity ion is obtained (this region includes about 30
copper ions). The dependence of the effective $g$-factor and the
ESR linewidth on the impurity concentration at low temperatures
indicates the interaction of clusters.

\end{abstract}

\pacs{PACS numbers:  75.10.Jm, 76.50.+g, 75.50.Ee}

]

\narrowtext

\section{INTRODUCTION}
Crystals of quasi-one-dimensional magnet CuGeO$_3$ exhibit
magnetic and crystallographic properties characteristic to
spin-Peierls magnets. \cite{Hase,Regnault,Nishi} The magnetic
structure of this compound is based on one-dimensional chains of
Cu$^{2+}$ ions ($S$ = 1/2) extended along the c-axis of the
orthorhombic crystal structure. \cite{structura} The value of the
exchange integral along these chains is 10.4 meV. \cite{Nishi}

Below the temperature of the spin-Peierls transition $T_{SP}$ =
14.5 K, the dimerization of chains occurs: i.e., magnetic ions
approach each other with the formation of pairs. The dimerization is
accompanied by the alternation of the exchange integral, which in turn
takes one of two values $J_{1,2}=J(1\pm\delta)$. An energy gap
$E\sim\delta J$ opens between the ground singlet state and triplet
excitations.  Due to the presence of the gap in the energy spectrum,
the magnetic susceptibility decreases, and the pure crystal without
defects becomes almost nonmagnetic at low temperatures. The
lattice transformation due to the dimerization is correlated in
space, and the dimers are located on a regular sublattice.

Interchain exchange interaction in CuGeO$_3$ is rather large (the
value of the exchange integral along the two directions orthogonal
to the chain is only by a factor of 10 and 100 less than the
exchange integral along the chains \cite{Regnault,Nishi}). For this
reason, in the absence of the spin-Peierls transition,
antiferromagnetic ordering had to be observed. However, the
spin-Peierls state is more preferable and is conserved down
to very low temperatures.

CuGeO$_3$ is the only inorganic spin-Peierls compound in which a
controlled substitution of magnetic ions is possible. The
introduction of impurities results in a local suppression of the
dimerization in the vicinity of the defect. As a result, the
temperature of the spin-Peierls transition decreases, and at low
temperatures the long-range antiferromagnetic order appears.
\cite{Zn&Ni-suzie,Mg&Zn-masuda,Ni-suzie,Glazkov,Si-doping}

The occurrence of the antiferromagnetic order and the suppression
of the dimerization order is explained in Refs.
\CITE{Fukuyama,Khomskii}. A cluster of antiferromagnetically
correlated spins is formed around the impurity ion. In the chain
of spins $S$ = 1/2 with alternating exchange interaction, the
antiferromagnetic correlations decay (see Ref. \CITE{corrlength})
thus forming wings of the cluster. As we recede from the defect,
the mean value of the  $z$-projection of the spin decreases
exponentially. Overlapping of the clusters' wings results in the
expansion of the region of antiferromagnetic correlations and in
the long-range antiferromagnetic order.

Substitution of the part of the Cu ions by Ni has two significant
differences compared to other dopants.

First, in the antiferromagnetically ordered phase, the easy axis
of anisotropy is directed along the $a$ axis, whereas for other
substituting impurities the easy axis of anisotropy is aligned
along $c$. \cite{Zn&Ni-suzie,Glazkov} Second, an anomalous
temperature dependence of the $g$-factor is observed in the
dimerized phase. At the deminishing of the temperature below the
transition point $T_{SP}$, the value of the effective $g$-factor
begins to decrease and achieves the value of 1.4 at low temperatures
for ${\bf H} \parallel c$. \cite{Glazkov} The anomalous value of the
$g$-factor can be explained by the existence of the antisymmetric
Dzyaloshinski--Moriya exchange interaction in the vicinity of the
defect. In a multispin system consisting of more than two spins, the
existence of the Dzyaloshinski--Moriya interaction along with the
symmetric Heisenberg  exchange  results in a strong
anisotropy of the effective $g$-factor and in the decrease of its
value.  \cite{Tsukerblat} Calculations based on the six-spin model
show that the existence of the antisymmetric exchange interaction with
the value of the exchange integral of about 30\% of the value of the
intrachain exchange interaction is sufficient for the description of
the  observed deviation  of the $g$-factor. \cite{Glazkov}

The present paper continues the study started in Ref.\CITE{Glazkov}.
Its purpose is to investigate high-quality samples of CuGeO$_3$
doped with nickel including those with a low content of the
impurity ($x < 1$ \%). The study of samples with a small
concentration of the impurity (when the average distance between
the impurity ions exceeds the characteristic cluster size) makes
it possible to observe the magnetic resonance of isolated
clusters. A noticeable difference of the $g$-factor of clusters
from the $g$-factor of excitations of the spin-Peierls matrix
makes it possible to differentiate between their ESR signals. In
turn, this fact opens the possibility to investigate the
interaction of clusters with the environment and between
themselves. The analysis of the experimental data allowed us to
determine the characteristic size of the cluster that is formed
around the impurity ion, namely, the size of the region where the
dimerization is destroyed and that of the region in which the
antisymmetric exchange interaction exists.

\section{ EXPERIMENTAL TECHNIQUE AND SAMPLES}

For the experiment, high-quality samples of
Cu$_{1-x}$Ni$_x$GeO$_3$ with the impurity concentration $x$ = 0.2\%
and $x$ = 0.8\% were grown. In order to analyze the dependence of
the $g$-factor on concentration, samples with higher
concentrations of the impurity ($x$ = 1.9\% and 3.0\%) were also
used.

To control the quality of the samples, a single crystal of the
pure compound grown following the same technology was used. At the
temperature of 4 K, the magnetic susceptibility of this sample
determined by the integral intensity of the ESR signal was about
4\% of the susceptibility at the transition temperature $T_{SP}$. This
value
corresponds to the residual concentration of the magnetic defects
per a copper ion equal to  0.05\%.

The investigations were performed at the frequency of 36 GHz and
the temperatures in the range 1.8--20 K with the help of an ESR
spectrometer with a transmission type cavity. The magnetic
resonance line was registered as the dependence of the intensity
of the microwave power transmitted through the resonator on the
magnetic field applied. In this case, the variation of the signal
is proportional to the imaginary part of the magnetic
susceptibility.

\section{ EXPERIMENTAL RESULTS}

At the  deminishing of the temperature  below the
spin-Peierls transition point (which is equal to 13.5 K for $x$ =
0.2\% and 12.0 K for $x$ = 0.8\%), the field of the resonance
absorption starts to increase. The temperature of the spin-Peierls
transition was determined from the beginning of the decrease of the
integral intensity of the ESR signal. The increase of the resonance
absorption field corresponds to the decrease of the $g$-factor.
The evolution of the ESR line with temperature is shown in Figs. 1
and 2.

The temperature dependences of the $g$-factor are presented in
Figs. 3 and 4. At low temperatures ($T < 4 K$), the values of the
$g$-factor remain constant and equal $g_a$ = 1.75, $g_b$ = 1.87,
and $g_c$ = 1.43 (for $x$ = 0.2\%).

For the sample with the impurity concentration $x$ = 0.2\%, the
magnetic resonance line splits into two components at the
temperature $T'$ = 7 K (Fig. 1). As the temperature decreases, one
of those components continues to move to the higher fields, and
its intensity increases. The second component remains in the field
close to the ESR field above $T_{SP}$, but its intensity decreases
and it almost vanishes as the temperature decreases further. The
width of the magnetic resonance line has its maximum at the
temperature close to the splitting temperature $T'$ (Fig. 5). A
similar splitting was observed at other orientations of the sample
with respect to the field for $x$ = 0.2\%; however, we were able
to follow it down to low temperatures only for ${\bf H}
\parallel c$ (this is due to the fact that for
this orientation there is the maximal difference of the
$g$-factors of two spectral components, which makes it possible to
distinguish the weak absorption line on the wing of the strong
one).

For the sample with the impurity concentration $x$ = 0.8\%, the
magnetic resonance line consists of the single component at all
temperatures; the maximum of the linewidth is observed in the
vicinity of $T'$ (Fig. 5).

In the paramagnetic phase the value of the $g$-factor is also
different from the value characteristic for the pure compound. The
dependence of the $g$-factor value on the impurity concentration
at $T > T_{SP}$ is shown in Fig. 6. As the impurity concentration
increases, the value of the $g$-factor decreases for all
orientations of the magnetic field.

For samples with the impurity concentration $x$ = 1.9\% and 3.0\%,
the long-range antiferromagnetic order takes place, the ordering
manifests itself in the transition from the gapless ESR
spectrum with a linear frequency-field dependence  to
a spectrum that is typical to antiferromagnets with orthorhombic
symmetry. The N\'{e}el temperatures are $T_N$= 2.5 K for $x$ = 1.9\%
and $T_N$= 3.5 K for $x$ = 3.0\%.

Comparison of the ESR lines at the minimal temperature (Fig. 7)
shows that the field of resonance absorption and the width of the
line of magnetic resonance are different for samples with
different concentration of impurity. Dependences of the linewidth
and the $g$-factor value on impurity concentration are presented
in Fig. 8. (For samples that exhibit the antiferromagnetic
ordering, the data were taken at $T = T_N$.) For small $x$, the
width of the ESR line linearly depends on the concentration.
Dependences of the ESR linewidth and of the $g$-factor value on
the orientation of the magnetic field at $T$ = 1.8 K for the
sample with $x$ = 0.8\% are shown in Fig. 9.

\section{DISCUSSION}

Before considering the quantitative analysis of the experimental
data, we will present a qualitative description.

According to the concept developed in
Refs.\CITE{Fukuyama,Khomskii}, a cluster of exchange- correlated
spins is formed around the impurity ion in the spin-Peierls matrix. Due
to the existence of the antisymmetric exchange interaction in this
cluster, the ESR of clusters is characterized by an unusually small
value of the $g$-factor $g_{cl}$. \cite{Glazkov} Clusters are
surrounded by a dimerized spin-Peierls matrix which is nonmagnetic in
the ground state.  Triplet excitations of the dimerized matrix are
characterized by the value of the $g$-factor of copper ions $g_{Cu}$
close to 2. Due to the exchange interaction of clusters with
excitations an ESR line with an intermediate value of the $g$-factor is
observed (the so- called exchange narrowing). At temperatures close to
the spin-Peierls transition temperature, when the concentration of
spin-Peierls excitations is large, an ESR line with the $g$-factor
close to the values characteristic to copper ions is observed. As the
temperature decreases, the concentration of triplet excitations
decreases due to the existence of an energy gap, and the ESR lime
shifts to the value characteristic of isolated clusters. As the
temperature decreases further, the effectiveness of the interaction of
clusters with excitations decreases, and the ESR line splits into two
components. A similar phenomenon was observed for the magnetic
resonance of temperature-activated spins in radicals \cite{Chestnut}.
At last, at low temperatures, when triplet excitations are practically
frozen out, the ESR line consists of two components --- a strong one,
characterized by the $g$-factor of clusters, and a weak one, which
represents the residual triplet excitations and magnetic defects. This
description corresponds to the observed evolution of the magnetic
resonance line for the samples with the impurity concentration of
0.2\%.

Similarly, one can explain the dependence of the $g$-factor on
impurity concentration at temperatures greater than the transition
temperature. In this case, one should consider the closest
neighborhood of the impurity ion in which antisymmetric
interaction exists as a cluster characterized by the anomalous
value of the $g$-factor $g_{cl}$. The ESR of copper ion chains is
characterized by the $g$-factor $g_{Cu}$. Due to the exchange
interaction of the cluster with the surrounding copper matrix, the
ESR line with an intermediate value of the $g$-factor is observed.
The more is the number of clusters, the more the magnetic
resonance line is shifted from $g_{Cu}$ to $g_{cl}$. Thus, the
value of the effective $g$-factor must decrease as the impurity
concentration increases.

In this reasoning, we assumed that the $g$-factor is the same for
all clusters. This is actually true if the interaction between
clusters is negligible. In this case the parameters of the ESR
line would depend only on the interaction of clusters with triplet
excitations. However, the fact, that the resonance absorption
fields for the samples with $x$ = 0.2\% and $x$ = 0.8\% differs,
shows that even for these impurity concentrations the interaction
between clusters must be taken into account. Clusters interact due
to the fact that their wings overlap \cite{Khomskii}; this makes
it possible to obtain a coarse evaluation of the cluster size
(assuming that clusters do not interact at $x$ = 0.2\%, and that
the interaction leads to the widening and shift of the line at $x$
= 0.8\%):

\begin{equation}
L\sim 1/0.008\sim 100.
\end{equation}

This result is overestimated since the distance between the
majority of clusters is less than the average one. Since
antiferromagnetic correlations at cluster wings are destroyed by
thermal fluctuations, the influence of the cluster interaction on
the ESR line should decrease with the increase of temperature.

We will assume that an isolated cluster is characterized by the
values of the $g$-factor observed for the sample with $x$ = 0.2\%
at the minimal temperature ($g$ = 1.75, $g$ = 1.87, and $g$ =
1.43). The values of the $g$-factor for excitations correspond to
the $g$-factor of copper ions in undistorted crystal environment,
i.e., in pure CuGeO$_3$ ($g$ = 2.15, $g$ = 2.26, and $g$ = 2.06,
and they are practically independent of temperature \cite{g_cu}).

In the subsequent analysis we use the following simplified model.
We assume that in the close neighborhood of the impurity ion of
size $L_{dim}$ the dimerization is suppressed and triplet
excitations of the spin-Peierls matrix do not reach this region.
Antiferromagnetic correlations exponentially decay with distance
from the defect. This attenuation is characterized by the magnetic
correlation length of dimerized chains
 $\xi\sim v/\Delta$ where $v$ is the speed of spin excitations and $\Delta$
 is the
energy gap, (see Ref.\CITE{corrlength}). In addition, there exists
the antisymmetric Dzyaloshinski--Moriya  exchange interaction in a
certain neighborhood of the impurity ion due to a local reduction
of symmetry. The size of this region is $L_{DM}<L_{dim}$. The
values of $L_{DM}$ and $L_{dim}$ are measured in interatomic
distances along the chains.

The analysis of the dependence of magnetic susceptibility on
temperature for a similar model was conducted in Refs.
\CITE{Grenier1,Grenier2}. The advantage of the ESR method is in
the fact that a noticeable difference in $g$-factors of clusters
and excitations makes it possible to separate their contributions.

\subsection {Interaction of Clusters with Excitations in the Molecular Field
Approximation}

First, we will consider the case $T < T_{SP}$ when spin chains are
dimerized. In the vicinity of the impurity ion, a cluster of
exchange-coupled spins with the total spin $S$ = 1/2 is formed. At
a large distance from the defect, the spin-Peierls matrix remains
unperturbed, and its magnetic properties are described by triplet
excitations, which are separated by a gap from the ground state.

The prolongation of antiferromagnetic correlations from the cluster
into the dimerized matrix results in the appearance of an
interaction between the cluster and excitations. Since this
interaction appears due to the exchange interaction between spins,
the average energy of the interaction can be written in the form

\begin{equation}
E_{int}=\sum_{i=1...n}J_{eff}(<{\bf S}_{cl}>\cdot <{\bf
S}_{Cu}(i)>)
\end{equation}

Here~~ $J_{eff}$~~~ is the~~ effective exchange integral,
$<{\bf S}_{cl}>$ is the average total spin value of the cluster, $<{\bf
S}_{Cu}>$ is the average spin value on the copper ion located away
from the cluster (it occurs due to triplet excitations). The
summation is performed over $n$ effective neighbors of the cluster
(since the major role is played by the interaction along spin
chains, we assume that $n$ = 2).

Following the molecular field theory, we obtain the following
system of self-consistent equations for average magnetization of a
cluster and a copper ion in the dimerized matrix:

\begin{eqnarray}
<\mu_{cl}>&=&\chi_{cl}^{(0)}(H+n\frac{J_{eff}}{g_{cl}g_{Cu}\mu_B^2}<\mu_{Cu}>),
\label{molfield}\\
<\mu_{Cu}>&=&\chi_{Cu}^{(0)}(H+\frac{J_{eff}}{g_{cl}g_{Cu}\mu_B^2}<\mu_{cl}>).
\nonumber \end{eqnarray}

Here $\chi_{cl,Cu}^{(0)}$ are susceptibilities per one cluster and
per one copper ion in the absence of the interaction.

From Eqs.(\ref{molfield}) one can derive following equations for
the magnetizations with regard for the interaction:

\begin{eqnarray}
\chi_{cl}&=&\chi_{cl}^{(0)}\frac{1+n\eta\chi_{Cu}^{(0)}}{1-n\eta^2\chi_{cl}^{(0)}\chi_{Cu}^{(0)}},
\label{chi}\\
\chi_{Cu}&=&\chi_{Cu}^{(0)}\frac{1+\eta\chi_{cl}^{(0)}}{1-n\eta^2\chi_{cl}^{(0)}\chi_{Cu}^{(0)}}
\nonumber
\end{eqnarray}

 where  $\eta=\frac{J_{eff}}{g_{Cu}g_{cl}\mu_B^2}$.

The magnetic susceptibility of a single isolated cluster obeys the
Curie law
\begin{equation}\label{Curie}
\chi_{cl}^{(0)}=\frac{g_{cl}^2\mu_B^2S(S+1)}{3kT}.
\end{equation}

For the susceptibility due to triplet excitations, we will use the
results obtained in Ref. \CITE{Grenier1,Grenier2}. In those studies,
an approximation of the magnetic susceptibility of pure CuGeO$_3$
crystals at temperatures below $T_{SP}$ was obtained
experimentally. This approximation of the molar susceptibility at
${\bf H} \parallel c$ has the form

 \begin{equation}
F(t)=(a_0+a_1t+a_2t^2)\exp(-\frac{A}{t}),~~~~~~~t=T/T_{SP},
\end{equation}

 where $a_0=26.0\times10^{-3}$ cgs units/mol, $a_1=-41.6\times10^{-3}$
 cgs units/mol, $a_2=28.2\times10^{-3}$ cgs units/mol, and $A = 2.39$.

Then, we have for the magnetic susceptibility per copper ion in
the dimerized matrix:
\begin{equation}
\label{chi_cu_0}
\chi_{Cu}^{(0)}=\left(\frac{g_{Cu}^{(i)}}{g_{Cu}^{(c)}}\right)^2\frac{F(T/T_{SP})}{N_A}.
\end{equation}
 Here  $g_{Cu}^{(i)}$ is the $g$-factor of the
copper ion in the corresponding direction.

If the impurity concentration is $x$, then the number of clusters
is $xN_A$ and the number of copper ions in the dimerized matrix is
$(1-xL_{dim})N_A$. Assuming that clusters do not interact, we
obtain the following formulas for the total susceptibility of
clusters and excitations:

\begin{eqnarray}
\tilde{\chi}_{cl}&=&xN_A\chi_{cl}\label{chi_dim},\\
\tilde{\chi}_{Cu}&=&(1-xL_{dim})N_A\chi_{Cu}.\nonumber
\end{eqnarray}

Equations (\ref{chi})-(\ref{chi_dim}) allow one to determine the
contribution of clusters and triplet excitations to the
susceptibility for all temperatures below the spin--Peierls
transition temperature. We will use this result later.

The case $T > T_{SP}$ can be treated in a similar way. As it has
already been mentioned above, in this case the neighborhood of an
impurity ion in which the Dzyaloshinski--Moriya exchange
interaction exists should be considered as a cluster; hence, the
characteristic size in Eq. (\ref{chi_dim}) is $L_{DM}$. Since the
susceptibility of spin chains weakly depends on temperature above
the transition temperature, we must set $T = T_{SP}$ in Eq.
\ref{chi_cu_0}.

\subsection{ Dependence of the $g$-factor on Temperature}

As it has already been mentioned above, the evolution of the ESR
line with the Ni concentration equal to 0.2\% (Fig. 1) has the
form typical of the exchange-narrowed two-component spectrum of
the magnetic resonance with the frequency of exchange jumps
dependent on temperature.

Following Refs.\CITE{Chestnut,Anderson}, we assume that the
influence of the exchange interaction on the magnetic resonance
spectrum of the system can be considered as random transitions
with the characteristic frequency $\omega_e$ between the states
with different Zeeman's frequencies
$\omega_2^{(0)}>\omega_1^{(0)}$.

The location of the center of gravity of the magnetic resonance
spectrum is independent of $\omega_e$ and is determined by the
formula

\begin{equation}\label{ave}
\overline{\omega}=\frac{\omega_1^{(0)}\tilde{\chi}_1+\omega_2^{(0)}\tilde{\chi}_2}{\tilde{\chi}_1+\tilde{\chi}_2},
\end{equation}

 where  $\tilde{\chi}_{1,2}$  are the susceptibilities of the
corresponding states with regard for the interaction between them.

Analysis of these random transitions by statistical methods (see
Ref.\CITE{Anderson}) show that the frequencies of the spectral
components and their widths are determined by the formulas

 \begin{eqnarray}\label{general}
\omega_{1,2}&=&\overline{\omega}+Im(\lambda_{1,2})\\
\Delta\omega_{1,2}&=&Re(\lambda_{1,2}),\nonumber
\end{eqnarray}

 where

 \begin{equation}
\lambda_{1,2}=\frac{1}{2}\{-[\omega_e-\imath\delta]\pm\sqrt{\omega_e^2-\Delta^2-2\imath\omega_e\delta}\},
\end{equation}

 \begin{equation} \label{gen_def}
\Delta=\omega_2^{(0)}-\omega_1^{(0)}~~~~~~~~\delta=\omega_1^{(0)}+\omega_2^{(0)}-2\overline{\omega}.
\end{equation}

 In the limit of  $\omega_e\gg\Delta$,
we have
\begin{eqnarray}
\omega_1=\overline{\omega}-\delta\frac{\delta^2+\Delta^2}{4\omega_e^2}&~~~~~~&\Delta\omega_1=\frac{\delta^2-\Delta^2}{4\omega_e}
\label{ultrafast}\\
\omega_2=\overline{\omega}+\delta&~~~~~~~&\Delta\omega_2=\omega_e.
\nonumber
\end{eqnarray}

Thus, the ESR spectrum consists of a narrow line close to
$\overline{\omega}$ and a wide background line.

In the absence of the interaction ($\omega_e=0$), we have
\begin{equation}
\omega_{1,2}=\omega_{1,2}^{(0)}~~~~~~~\Delta\omega_{1,2}=0
\end{equation}
 which corresponds to two narrow spectral components at the
frequencies  $\omega_1^{(0)}$ and $\omega_2^{(0)}$.

Qualitatively, this corresponds to the observed transition from
the ESR line consisting of a single component to the two-component
line. In this model, we neglect the intrinsic widths of lines in
both states of the system.

Magnetic properties of the doped spin-Peierls system at $T > T_N$
correspond to free spins of clusters and triplet excitations of
the dimerized matrix. The difference in $g$-factors of clusters
and excitations lead to differences in Zeeman's frequencies.

The presence of an energy gap leads to a dependence of the
concentration of triplet excitations on temperature. In this case,
the frequency of exchange jumps $\omega_e$ also depends on
temperature as

\begin{equation}\label{omega_e}
\omega_{e}(t)=\Omega_{e}\exp\{-\frac{E(t)/T_{SP}}{t}\},~~~~~~~t=T/T_{SP}.
\end{equation}

The dependence of the energy gap on temperature can be
approximated as follows (see Refs. \CITE{Ni-suzie,Martin}):

\begin{equation}\label{gap(T)}
E(t)=E(0)(1-t)^a,~~~~~ a\approx0.1,~ t=T/T_{SP}.
\end{equation}

The magnitude of the energy gap at $T$ = 0 K is related to the
transition temperature by the equation (see Ref. \CITE{Pytte})
\begin{equation}\label{gap0}
E(0)=1.76kT_{SP}.
\end{equation}

Equations (\ref{general}-\ref{gen_def}), (\ref{ave}) and
(\ref{omega_e})-(\ref{gap0}) make it possible to obtain
temperature dependences of the resonance absorption frequencies
($g$-factors) and widths of spectrum components. To take into
account the interaction of clusters with triplet excitations, we
use the molecular field approximation (\ref{chi})-(\ref{chi_dim}).

The temperature dependences of the $g$-factor and the width of the
magnetic resonance line are described with the help of three
fitting parameters -- the size of the region of suppressed
dimerization $L_{dim}$, the effective exchange integral value
$J_{eff}$, and the preexponential coefficient of the exchange
frequency $\Omega_e$.

This model assumes that clusters do not directly interact. As it has
been mentioned above, the influence of interaction of the clusters
decreases with increasing temperature. For this reason, when
choosing the parameters, we used the temperature dependence of
the $g$-factor at $T > T'$ = 7 K for all basic orientations of
both samples and the temperature dependence of the $g$-factor
below $T'$ for the sample with the impurity content 0.2\% for
${\bf H} \parallel c$.

Thus, the following values of the fitting parameters were
obtained:

\begin{eqnarray}\label{first_res}
L_{dim}=32\pm2,~~J_{eff}=-(13\pm1)K,~ \nonumber\\
\Omega_e=(2.2\pm0.3)\times
10^{12} sec^{-1},
 \end{eqnarray}

Note that  $\hbar\Omega_e/k\sim$16 K, which is close to $J_{eff}$.
This result could be expected since $\Omega_e$ and $J_{eff}$ must
be determined by the magnitude of the intrachain exchange
integral.

The comparison of the theoretical and experimental results is
illustrated in Figs. 3, 4. The theoretical dependences provide an
accurate description of the experimental data for the sample with
the impurity concentration 0.2\%; however, for the sample with $x$
= 0.8\%, there is a disagreement at low temperatures, which we
attribute to the interaction of clusters.

The value of the suppressed dimerization region obtained here
coincides with the result of Ref. \CITE{Grenier2} which was
obtained by the analysis of static susceptibilities.

We also can make a coarse evaluation of the impurity concentration
at which the long-range spin-Peierls order must be completely
destroyed: $x_C$ = 1/$L_{dim} \sim 0.03$, which is in good
agreement with the result obtained in Ref.\CITE{Koide}.

\subsection{ Dependence of the Width of the Magnetic
Resonance Line on Temperature}

On the basis of the model described above, we can derive the
dependence of the width of the ESR line on temperature. The
comparison with experimental data is presented in Fig. 5. For
convenience, the width of the line at the spin-Peierls transition
point is added to the theoretical dependences. No additional
adjustable parameters were used.

For the sample with the impurity concentration 0.2\%, the
agreement of the theory with the experiment is very good. The
theory provides correct location of the maximum of the linewidth
and correct value of it at this point. The best agreement between the
theory and the experiment is achieved for ${\bf H} \parallel c$.
This could be expected, since in this case one of the basic
assumption of this model is best satisfied: namely, that the
intrinsic linewidth of the spectral components can be neglected as
compared with the splitting between them.

For the sample with $x$ = 0.8\%, a disagreement of the theory and
the experiment is observed. The location of the maximum is
determined rather well; however, the behavior of the linewidth at
low temperatures is different from that predicted by the model. We
attribute this fact to interaction between clusters.

\subsection {Dependence of the $g$-factor on Concentration
above the Temperature of the Spin-Peierls Transition}

Approach developed above can be also applied to the description of
the dependence of the $g$-factor value on impurity concentration
above  $T_{SP}$. In this case, we consider as a cluster the
neighborhood of the impurity ion of size $L_{DM}$ where the
Dzyaloshinski--Moriya exchange interaction exists.

At temperatures close to $T_{SP}$, the condition
 $\omega_e\gg\Delta$
($\omega_e\sim\Omega_e\sim10^{12} sec^{-1}$, $\Delta\sim10^{10}
sec^{-1}$) holds. Hence, simplified Eqs. (\ref{ultrafast}) can be
used. Neglecting the terms of order $\Delta^2/\omega_e^2$, we
obtain the following equation for the mean value of the $g$-factor
(this equation is similar to (\ref{ave}):

\begin{equation}\label{g_ave}
\overline{g}=\frac{g_{cl}\tilde{\chi}_{cl}+g_{Cu}\tilde{\chi}_{Cu}}{\tilde{\chi}_{cl}+\tilde{\chi}_{Cu}}.
\end{equation}

As before (see Eqs. (\ref{chi})-\ref{chi_dim}), susceptibilities
$\tilde{\chi}$ are determined in the molecular field
approximation. We assume that in the absence of interaction, the
cluster susceptibility is described by the Curie law
(\ref{Curie}), and the susceptibility of the copper ions
surrounding the cluster is independent of temperature and equals
the susceptibility at the point of the spin-Peierls transition
 (\ref{Curie}).

We do not present the expression for the dependence of the
$g$-factor value on concentration because it is too cumbersome.
This expression includes two parameters: the effective exchange
integral $J_{eff}$ and $L_{DM}$. The parameter $J_{eff}$ has
already been determined earlier. This leaves us a single
adjustable parameter to describe the dependence of the $g$-factor
on impurity concentration for all orientations of the sample with
respect to the magnetic field. As it has already been mentioned,
the value of the impurity concentration at which clusters can be
considered as uninteracting increases with temperature. Thus, at
high temperatures our approach can be applied even in the case of
large concentrations. Figure 6 presents data for samples with
nickel concentrations up to 3.3\% at the temperature of 15 K along
with theoretical curves. The experimental dependences correspond
to $L_{DM} =18 \pm 2$.

\subsection{Dependence of the ESR Linewidth on Concentration
at Low Temperatures. Interaction of Clusters}

The difference of the ESR lines of samples with the impurity
concentration 0.2\% and 0.8\% (Fig. 7) indicates that clusters
interact. The dependence of the linewidth on impurity
concentration for small $x$ is linear (Fig. 8). A linear dependence
of the ESR linewidth on the concentration of magnetic centers was
observed in experiments with diluted paramagnets (paramagnetic
centers in a diamagnetic crystal) (see, e.g., Ref.\CITE{Altshuler}).

As a possible cause of the observed linewidth one can suggest
long-range dipole--dipole interactions or exchange interactions
occuring due to overlapping of wings of nearby clusters.

In order to estimate the contribution of the dipole--dipole
interaction to the linewidth, we notice that the dipole field
magnitude is about  10 Oe at the distance of 10$\AA$ from the
magnetic moment $\mu_B$. Thus, the observed ESR linewidth $\sim$ 1
kOe cannot be explained by the existence of the dipole--dipole
interaction between impurity ions.

Therefore, the linewidth must be determined by the antisymmetric
or anisotropic exchange interaction of clusters. Reorientation of
clusters due to thermal fluctuations leads to the appearance of a
random effective field $H_{eff}$, which determines the linewidth.

Due to the random distribution of the impurities, the number of
closely-spaced clusters constitutes a noticeable part of the total
number of clusters. In the one-dimensional case, the probability
to have an impurity ion at the distance of $n$ interatomic
distances from the given ion at the impurity concentration $x$ is
\begin{equation}\label{prob}
p(n)=x(1-x)^n.
\end{equation}
Then, the probability that the distance between impurity ions is
less than $N$ is
\begin{equation}
P(n<N)=\sum_{n=0}^{N-1}p(n)=1-(1-x)^N.
\end{equation}

In the limiting case $x\ll 1$ we obtain $P(n<N)\approx Nx$.
Thus, if the impurity concentration is 1\% (and the average
distance between impurity ions in a chain is 100 interatomic
distances) the part of the clusters that are closer to each other
than $L_{dim}$ = 32 is about of 30\%.

On the basis of the observed values of the width of the magnetic
resonance line, we can give a coarse evaluation of the magnitude
of the random effective magnetic field generated by clusters.

Since antiferromagnetic correlations decrease exponentially when
moving away from the defect into the dimerized matrix, we assume
that the average value of the effective field depends on the
distance $L$ from the region of the destroyed dimerization
according to the law
\begin{equation}
H_{eff}=H_0\exp(-L/\xi),
\end{equation}
 where $\xi$  is the magnetic correlation length and $H_0$ is the
effective field on the boundary of the suppressed dimerization
region.

Averaging over $L$ with the help of distribution  (\ref{prob}) and
taking into account that $x$ is small, we obtain the following
estimate for the width of the ESR line:
\begin{equation}
\Delta H\sim\sum_{L=0}^\infty
H_{eff}(L)p(L)=\frac{xH_0}{1-(1-x)\exp(-1/\xi)}\approx\xi xH_0.
\end{equation}

Hence, setting $\xi\approx 10$ (see Ref.\CITE{Kojima}) and taking
into account that at $x\sim1\%$ the linewidth $\Delta H\sim 1$ kOe
(see Fig. 8), we obtain for $H_0$ the estimate $H_0 \sim 10$ kOe.
Such a magnitude of the effective field corresponds to the energy
of order of 1 K, which is about 1\% of the intrachain exchange
integral.

Additional information on the nature of the interaction that
determines the width of the magnetic resonance line can be
obtained with the help of the angular dependence of the linewidth.
The dependences of the effective $g$-factor and the ESR linewidth
on the orientation of the magnetic field in the plane $bc$ of the
crystal with $x$ = 0.8\% are presented in Fig. 9. The angular
dependence of the $g$-factor is accurately approximated by the
function
\begin{equation}\label{g_angul}
g_{eff}^2=g_c^2\cos^2\phi+g_b^2\sin^2\phi,
\end{equation}
where $\phi$ is the angle in the plane $bc$ measured from the axis
$c$. Thus, the anisotropy of the $g$-factor can be described in
terms of the principal values of the $g$-tensor.

The contribution of the antisymmetric exchange interaction to the
angular dependence of the linewidth is $\pi$-periodic, and the
contribution of the anisotropic symmetric interaction is
$\pi/2$-periodic. \cite{Lohman} In the case under consideration,
both contributions are present. Figure 9 illustrates fitting of
experimental data for the ESR linewidth by the function
$A+B\cos(2\phi)+C\cos(4\phi)$. However, one must take into account
the fact that the anisotropy of the $g$- factor also affects the
angular dependence of the linewidth, and this influence is
periodic with the period equal to that of the angular dependence
of the $g$-factor (\ref{g_angul}), i.e., $\pi$.

The magnitude of the parameter $D$ of the anisotropic symmetric
exchange is related to the isotropic exchange integral $J$ by the
equation (see Ref.\CITE{Yablokov})
\begin{equation}
D\sim\left(\frac{\Delta g}{g}\right)^2J,
\end{equation}
where $\Delta g = g - 2$. In CuGeO$_3$, $(\Delta g/g) \sim 0.1$,
which yields an estimate of 1 K for $D$. Thus, it is possible that
the observed magnitude of the ESR line is explained by the
existence of the symmetric anisotropic exchange interaction.

\section{CONCLUSIONS}

When studying high-quality samples of the spin-Peierls magnet
CuGeO$_3$ doped with nickel with a small impurity concentration
$x < $1\%, it was found that the $g$-factor decreases with
temperature to unusually small values (down to 1.4). This fact is
due to formation of the clusters of antiferromagnetically
correlated spins with antisymmetric exchange interaction around
impurity ions. Above the transition temperature, the value of the
$g$-factor decreases as the impurity concentration increases.

The dependence of the $g$-factor on temperature and concentration
can be explained in the framework of the model of exchange
narrowing. An analysis of data allows one to evaluate the size of
the region around an impurity in which the dimerization is
suppressed ($L_{dim} \approx 30$ interatomic distances) and the
size of the region in which the antisymmetric exchange interaction
exists ($L_{DM} \approx   20$ interatomic distances).

Experimental data show that even at small impurity concentrations,
the interaction between clusters plays an important role at low
temperatures. The magnitude and the angular dependence of the
width of the magnetic resonance line suggest the existence of an
anisotropic exchange interaction in CuGeO$_3$.

\section {ACKNOWLEDGMENTS}

This work was supported by the Russian Foundation for Basic
Research and Deutsche Forschungsgemeinschaft (the joint project
No. 01-02-04007); INTAS, project No. 99-0155; the U.S. Civilian
Research and Development Foundation (CRDF), project No. RP1-2097;
and the BRHE Foundation, project No. REC007. The authors are
grateful to M.V. Eremin for his interest to the work and useful
discussions. One of the authors (V. Glazkov) is also grateful to
the Forshungcentrum Julih Gmbh for the continuous support of his
research work.

\section{Figure captions}
\begin{itemize}

\item[Fig.1]   The temperature evolution of the ESR line at $x$ = 0.2\%,
${\bf H} \parallel c$, and $f$ = 36 GHz. The vertical segments in
figures $a$ and $b$ correspond to the same amplitude of the
signal.

\item[Fig.2]  The temperature evolution of the ESR line at $x$ = 0.8\%,
${\bf H} \parallel c$, and $f$ = 36 GHz.

\item[Fig.3] Dependence of the effective $g$-factor on temperature for
the sample with the impurity content $x$ = 0.2\%:
 $\bullet$ - {\bf H}$\parallel a$,
$\Box$ - {\bf H}$\parallel b$, $\bigtriangledown$ - {\bf
H}$\parallel c$. Solid curves correspond to the theoretical
calculation.

\item[Fig.4]  Dependence of the effective $g$-factor on temperature for
the sample with the impurity content $x$ = 0.8\%:
 $\bullet$ - {\bf H}$\parallel a$,
$\Box$ - {\bf H}$\parallel b$, $\bigtriangledown$ - {\bf
H}$\parallel c$. Solid curves correspond to the theoretical
calculation.

\item[Fig.5]  Dependence of the half-width of the ESR line on
temperature for $x$=0.2\% (a) and $x$=0.8\% (b). {\bf H}$\parallel
c$, $f=$36~GHz . Solid curves correspond to the theoretical
calculation.

\item[Fig.6]
Dependence of the effective $g$-factor on impurity
concentration at $T$ = 15 K:
 $\bigcirc$ - {\bf H}$\parallel a$,
$\Box$ - {\bf H}$\parallel b$, $\bigtriangledown$ - {\bf
H}$\parallel c$. Solid curves correspond to the theoretical
calculation, and black symbols correspond to the data of the study
\cite{Glazkov}.

\item[Fig.7]  Comparison of the ESR lines for $x$ = 0.2\% and $x$ =
0.8\% at $T$ = 1.8 K,
 H$\parallel c$, and $f$ = 36 GHz.

\item[Fig.8] Dependence of the half-width of the ESR line on impurity
concentration for H$\parallel c$: $T = T_N$ = 2.5 K for $x$ =
1.9\%, $T = T_N$ = 3.5 K for $x$ = 3.0\%, and $T$ = 1.8 K for $x$
= 0.2\% and $x$ = 0.9\%.

\item[Fig.9] Angular dependence of the width of the magnetic resonance
line ($\bigcirc$) and the effective $g$-factor ($\Box$) for the
field applied in the plane $bc$. Solid curves correspond to the
dependence $A+B\cos(2\phi)+C\cos(4\phi)$,  $x$ = 0.8\%, $T$ = 1.8
K, and $f$ = 36 GHz; $\phi = 0^o$ corresponds to ${\bf H}\parallel
c$.

\end{itemize}

\end{document}